\documentclass[preprint]{aastex6}
\usepackage{amssymb,amsmath}
\usepackage{booktabs}
\usepackage{multirow}
\usepackage{longtable}
\usepackage{arydshln}
\usepackage{colortbl}

\begin{document}

\slugcomment{Accepted to AJ}

\title{Col-OSSOS: $z$ Band Photometry Reveals Three Distinct TNO Surface Types}

\author{Rosemary E. Pike\altaffilmark{1}}
\author{Wesley C. Fraser\altaffilmark{2}}
\author{Megan E. Schwamb\altaffilmark{3}}
\author{J.  J.  Kavelaars\altaffilmark{4,5}}
\author{Michael Marsset\altaffilmark{2}}
\author{Michele T. Bannister\altaffilmark{2}}
\author{Matthew J. Lehner\altaffilmark{1,6,7}} 
\author{Shiang-Yu Wang\altaffilmark{1}} 
\author{Mike Alexandersen\altaffilmark{1}}
\author{Ying-Tung Chen\altaffilmark{1}}
\author{Brett J. Gladman\altaffilmark{8}}
\author{Stephen Gwyn\altaffilmark{9}}
\author{Jean-Marc Petit\altaffilmark{10}}
\author{Kathryn Volk\altaffilmark{11}}

\altaffiltext{1}{Institute of Astronomy and Astrophysics, Academia Sinica, National Taiwan University, Taipei, Taiwan}
\altaffiltext{2}{Astrophysics Research Centre, Queen's University, Belfast, Belfast, UK, BT7 1NN}
\altaffiltext{3}{Gemini Observatory, Northern Operations Center, 670 North A'ohoku Place, Hilo, HI 96720, USA}
\altaffiltext{4}{Department of Physics and Astronomy, University of Victoria, Victoria, BC, V8P 5C2, Canada}
\altaffiltext{5}{NRC-Herzberg Astronomy and Astrophysics, National Research Council of Canada, Victoria, British Columbia, Canada}
\altaffiltext{6}{Department of Physics and Astronomy, University of Pennsylvania, 209 South 33rd Street, Philadelphia, PA 19104, USA}
\altaffiltext{7}{Harvard-Smithsonian Center for Astrophysics, 60 Garden Street, Cambridge, MA 02138, USA}
\altaffiltext{8}{Department of Physics and Astronomy, University of British Columbia, Vancouver, BC, V6T 1Z1, Canada}
\altaffiltext{9}{Canadian Astronomy Data Centre, 5071 West Saanich Road, Victoria, BC, V9E 2E7, Canada}
\altaffiltext{10}{Institut UTINAM UMR6213, CNRS, Univ. Bourgogne Franche-Comt\'{e}, OSU Theta F-25000 Besan\c{c}on, France}
\altaffiltext{11}{Lunar and Planetary Laboratory, University of Arizona, 1629 E University Blvd, Tucson, AZ 85721, United States}

\begin{abstract}
Several different classes of trans-Neptunian objects (TNOs) have been identified based on their optical and near-infrared colors.
As part of the Colours of the Outer Solar System Origins Survey, we have obtained $g$, $r$, and $z$ band photometry of 26 TNOs using Subaru and Gemini Observatories.
Previous color surveys have not utilized $z$ band reflectance, and the inclusion of this band reveals significant surface reflectance variations between sub-populations.
The colors of TNOs in $g-r$ and $r-z$ show obvious structure, and appear consistent with the previously measured bi-modality in $g-r$.
The distribution of colors of the two dynamically excited surface types can be modeled using the two-component mixing models from \cite{fraser2012}.
With the combination of $g-r$ and $r-z$, the dynamically excited classes can be separated cleanly into red and neutral surface classes.
In $g - r$ and $r - z$, the two dynamically excited surface groups are also clearly distinct from the cold classical TNO surfaces, which are red, with $g-r\gtrsim$0.85 and $r-z\lesssim$0.6, while all dynamically excited objects with similar $g-r$ colors exhibit redder $r-z$ colors. 
The $z$ band photometry makes it possible for the first time to differentiate the red excited TNO surfaces from the red cold classical TNO surfaces.
The discovery of different $r-z$ colors for these cold classical TNOs makes it possible to search for cold classical surfaces in other regions of the Kuiper belt and to completely separate cold classical TNOs from the dynamically excited population, which overlaps in orbital parameter space.

\end{abstract}


\section{Introduction}

Trans-Neptunian objects (TNOs) in the outer Solar System exhibit a broad range of surface properties.
The vast majority of TNOs are too faint for spectroscopic studies, so broad band surface reflectance is used to provide constraints on surface composition.
TNOs, in general, have red optical colors in $g$, $r$, and $i$ bands; even the `neutral' TNO surfaces, sometimes referred to as `blue' in the literature, are slightly redder than Solar.
It is well accepted that the small dynamically excited TNOs and Centaurs exhibit a bimodal color distribution, with red and neutral classes \citep[e.g.][]{tegler1998,peixinho2003,peixinho,peixinho2015,tegler03,tegler2016,fraser2012,wong2017}.

Color surveys of TNOs in more than two bands have revealed additional complexities and correlations in the surface reflectance of these objects.
Observations in $i$ band correlate strongly with $g$ and $r$ \citep{Sheppard2012,ofek2012}, suggesting that across the $g$ to $i$ wavelength range the same spectral feature is being probed.
The dynamically excited TNOs also show correlations between the optical and near-infrared colors \citep{fraser2012}.
The observed correlations in color have revealed different surface classes, though the exact number of classes is debated in the literature \citep{barucci2005,dalleore2013,fraser2012}.

TNOs are often subdivided based on their dynamical classifications into cold classical TNOs and dynamically excited objects, including scattering, detached, hot classical, and resonant TNOs \citep{brown2001,gladman2008}.
A range of criteria are used to identify cold classical TNOs in the literature; small eccentricities, semi-major axes between the 3:2 and 2:1 Neptune resonances, and an inclination cut at 4--7$^{\circ}$ typically identifies cold classical objects with minimal contamination from the hot population.
Most cold classical TNOs exhibit similar red colors to the red excited TNOs, with the color distributions of both classes occupying similar ranges in $B-R$, $g-r$, and in the NIR \citep[e.g.][]{tegler03,fraser2012,colossos}.
The red cold classicals, however, exhibit higher albedos \citep{brucker2009,vilenius2014} than the dynamically excited red objects, implying that they occupy a different compositional class.

A number of different TNO taxonomies have been proposed.
\cite{barucci2005} use the colors derived from $BVRIJ$ photometry to classify TNOs into four different classes using the G-mode statistical analysis method for classifying asteroids.
A different technique that utilized a modified K-means clustering technique applied to multi-band optical photometry and optical albedos found 10 surface types \citep{dalleore2013}.
\cite{fraser2012} and \cite{fraser2015} found that the small dynamically excited TNOs fall into only two surface classes which exhibit a range of optical colors.
The true number of TNO compositional classes remains an open question.

Extending photometric surveys of TNOs into the previously unexplored $z$ band provides new insight into TNO surface classifications.
In part due to detector sensitivity, $z$ band photometry has not been utilized as a tool for probing TNO surface types.
This wavelength range may be sensitive to the presence or absence of organics and silicates on minor planet surfaces.
Here we present $g$, $r$, and $z$ band photometry of 26 TNOs.
We find three distinct TNO surface types that result from classifying based on these surface colors.

\section{Sample Selection}

The targets for this work are TNOs found by two large surveys.
23 targets are from the Outer Solar System Origins Survey \citep[OSSOS,][]{bannister2016}, a large TNO discovery survey executed on the Canada-France Hawaii Telescope (CFHT) from 2013--2017.
Three additional targets, the full known sample of 5:1 resonators \citep{pike2015}, are included from the Canada-France Ecliptic Plane Survey \citep[CFEPS,][]{cfeps,hilat2016}.
All 26 targets from these two surveys, listed in Table \ref{targets}, are from a flux-limited sub-sample of these two surveys, with $r<23.6$.

\begin{table}[h!]
\setlength{\tabcolsep}{2.5pt}
\small
\caption{Barycentric Orbit Parameters and Discovery Characteristics of Targets}
\label{targets}
\begin{center}
\begin{tabular}{ l c  c  c  c  c  c  c c  c c c c }
Survey &   MPC & Class & $a$ & $e$ &  $i$ &   Discovery & H${_r}$ (H${_g}$) &  $r$ \\
 ID & ID& & (AU)& &  (degrees) &  $r$$^{\prime}$($g$$^{\prime}$) mag &   mag &  (AU) \\ \hline
 HL7j4 & 2007 LF$_{38}$& res 5:1 & 87.57 $\pm$ 0.03 & 0.56 & 35.83  & 22.53 $\pm$ 0.09 & 5.5 &48.4 \\
o3l79 &  2013 SA$_{100}$ & hc & 46.30 $\pm$ 0.01 & 0.17 & 8.48  & 22.79 $\pm$ 0.04 & 5.75 & 50.54  \\
o4h50 &  2014 UE$_{225}$ & cc & 43.71 $\pm$ 0.00 & 0.07 & 4.49  & 22.68 $\pm$ 0.04 & 6.00 & 46.56  \\
o3l77 &  2013 UQ$_{15}$ & hc & 42.77 $\pm$ 0.01 & 0.11 & 27.34  & 22.96 $\pm$ 0.17 & 6.10 & 47.53  \\
o3l76 &  2013 SQ$_{99}$ & cc & 44.15 $\pm$ 0.01 & 0.09 & 3.47  & 23.12 $\pm$ 0.06 & 6.37 & 47.30  \\
o5t31  &  2015 RT$_{245}$ & cc & 44.39 $\pm$ 0.03 & 0.08 & 0.96  & 22.87 $\pm$ 0.04 & 6.57 & 41.89  \\
o3l39 &  2016 BP$_{81}$ & cc(bb) & 43.67 $\pm$ 0.01 & 0.08 & 4.18  & 22.96 $\pm$ 0.06 & 6.59 & 42.48  \\
o3l43 &  2013 UL$_{15}$ & cc & 45.78 $\pm$ 0.02 & 0.10 & 2.02  & 23.02 $\pm$ 0.11 & 6.59 & 43.04  \\
L3y02 &   2003 YQ$_{179}$& res 5:1 & 88.41 $\pm$ 0.02 & 0.58 & 20.87  & (23.38 $\pm$ 0.09) & (7.3) & 39.3 \\
o4h45 &  2014 UD$_{225}$ & cc(bb) & 43.37 $\pm$ 0.01 & 0.13 & 3.66  & 23.07 $\pm$ 0.05 & 6.61 & 44.31  \\
o5t09PD & 2014 UA$_{225}$ & det & 67.76 $\pm$ 0.01 & 0.46 & 3.58  & 22.50 $\pm$ 0.02 & 6.74 & 36.76  \\
o3l63  & 2013 UN$_{15}$ & cc & 45.14 $\pm$ 0.01 & 0.06 & 3.36  & 23.63 $\pm$ 0.21 & 7.01 & 45.10  \\
o3l46  & 2013 UP$_{15}$ & cc & 46.62 $\pm$ 0.00 & 0.08 & 2.47  & 23.61 $\pm$ 0.09 & 7.15 & 43.42  \\
o4h20  & 2014 UL$_{225}$ & hc & 46.35 $\pm$ 0.01 & 0.20 & 7.95  & 22.97 $\pm$ 0.06 & 7.18 & 37.95  \\
HL7c1  & 2007 FN$_{51}$& res 5:1  &  87.49 $\pm$ 0.07 & 0.62 & 23.24  & 23.20 $\pm$ 0.06 & 7.2 &39.1 \\
o4h31 & 2014 UM$_{225}$ & res 9:5 & 44.48 $\pm$ 0.00 & 0.01 & 18.29  & 23.26 $\pm$ 0.09 & 7.22 & 40.15  \\
o4h29 & 2014 UH$_{225}$ &  hc & 38.64 $\pm$ 0.00 & 0.04 & 29.53  & 23.33 $\pm$ 0.07 & 7.32 & 40.06  \\
o5t11PD   & 2001 QE$_{298}$ & res 7:4 & 43.71 $\pm$ 0.00 & 0.16 & 3.66  & 23.17 $\pm$ 0.04 & 7.38 & 36.97  \\
o5s16PD & 2004 PB$_{112}$ & res 27:4 & 107.52 $\pm$ 0.02 & 0.67 & 15.43  & 22.99 $\pm$ 0.03 & 7.39 & 35.51  \\
o4h19  & 2014 UK$_{225}$ & hc & 43.52 $\pm$ 0.03 & 0.13 & 10.69  & 23.20 $\pm$ 0.06 & 7.40 & 38.08  \\
o3l15  & 2013 SZ$_{99}$ & hc & 38.28 $\pm$ 0.00 & 0.02 & 19.84  & 23.41 $\pm$ 0.17 & 7.52 & 38.74  \\
o3l06PD  & 2001 QF$_{331}$ & res 5:3 & 42.25 $\pm$ 0.02 & 0.25 & 2.67  & 22.69 $\pm$ 0.07 & 7.54 & 32.73  \\
o3l09  & 2013 US$_{15}$ & res 4:3 & 36.38 $\pm$ 0.01 & 0.07 & 2.02  & 23.22 $\pm$ 0.16 & 7.76 & 34.45  \\
o5s06   & 2015 RW$_{245}$ & sca & 56.47 $\pm$ 0.02 & 0.53 & 13.30  & 22.90 $\pm$ 0.03 & 8.53 & 26.58  \\
o5t04   & 2015 RU$_{245}$ & sca & 30.99 $\pm$ 0.01 & 0.29 & 13.75  & 22.99 $\pm$ 0.04 & 9.32 & 22.72  \\
o5s05   & 2015 RV$_{245}$ & cen & 21.98 $\pm$ 0.01 & 0.48 & 15.39  & 23.21 $\pm$ 0.04 & 10.10 & 19.89  \\
o4h01 & 2014 UJ$_{225}$ &  cen & 23.18 $\pm$ 0.01 & 0.38 & 21.32  & 22.71 $\pm$ 0.10 & 10.26 & 17.76  \\
o3l01  & 2013 UR$_{15}$ & sca & 55.82 $\pm$ 0.03 & 0.72 & 22.25  & 23.06 $\pm$ 0.07 & 10.89 & 16.05  \\
\hline
\end{tabular}
\end{center}
\footnotesize{Note.  Dynamical classifications are based in precision OSSOS \citep{bannister2016} and CFEPS \citep{cfeps,hilat2016} astrometry, via 10~Myr integrations of the best-fit and extremal-fit orbits from \cite{bernstein}-- cc: cold classical;  cc(bb): blue binary cold classical \citep{fraser2017};  hc: hot classical; res: resonant; sca: scattering; cen: centaur; det: detached\\ Columns include: semi-major axis $a$, eccentricity $e$, inclination $i$, $r$$^{\prime}$ or $g$$^{\prime}$ magnitude, Solar System absolute magnitude $H$, and distance at discovery $r$.  All digits quoted for $e$, $i$, and $r$ are significant. \\ Survey IDs beginning with `o' indicate the TNO is an OSSOS object.  The object with a survey ID beginning with `L' is from the ecliptic portion of CFEPS \citep{cfeps}, and the objects with IDs beginning with `H' are from the high-latitude component of CFEPS  \citep{hilat2016}.}
\end{table}

\section{Photometry}
\subsection{Observations}

Two programs were used to gather colors of our targets.
The OSSOS targets were measured by the Colours of the Outer Solar System Origins Survey (Col-OSSOS) Large Program on Gemini North (GN-2014B-LP-1, GN-2015A-LP-1, GN-2015B-LP-1, GN-2016B-LP-1; Principal Investigator Wesley Fraser), which obtains near-simultaneous $g$, $r$, and $J$ band photometry of a flux limited subset of the OSSOS TNO sample, $r<23.6$ \citep{colossos}.
The Col-OSSOS project began in August 2014 and aims to obtain photometry of $>$100 TNOs with better than 5\% precision in all bands over several years of observations.
Photometry in $g$ and $r$ band was acquired with the Gemini Multi-Object Spectrograph \citep[GMOS,][]{hook2004}.
A random sub-sample of the Col-OSSOS targets were observed in $z$ band as well, through extensions to Col-OSSOS utilizing GMOS, or through observations with Subaru Suprime-Cam \citep{miyazaki2002} which include some combination of $Rc$, $i$, and $z$ band images.
The CFEPS \citep{cfeps,hilat2016} targets were measured in a separate Gemini Observatory Fast Turnaround program (GN-2015B-FT-28; Principal Investigator Rosemary Pike) using GMOS in $g$, $r$, and $z$ band.
This program has similar color measurement uncertainty to Col-OSSOS.
Here we focus on the $g$, $r$, and $z$ band photometry from these two observing programs.

Data acquisition required excellent sky conditions for accurate photometry.
All photometry was taken above an airmass of 2 in dark/gray time.
All data were acquired in photometric conditions and with a seeing of `IQ70' or better, which corresponds to a delivered full width half maximum of $<1.2\arcsec$ at an airmass of 2.
The delivered image quality ranged from 0.29--0.99$\arcsec$, and the median seeing was 0.53$\arcsec$, much better than the minimum requirements.

The Gemini and Subaru data were prepared for analysis using standard data reduction packages.
Standard debias and flat fielding of the Col-OSSOS \citep{colossos} and Fast Turnaround data from GMOS were performed using the Gemini-IRAF package.
GMOS has a field of view of 330$\arcsec\times330\arcsec$ and a pixel scale of 0.0728$\arcsec$ with 1$\times$1 binning.
Subaru data were acquired in $Rc$, $i$, and $z$ band with Suprime-Cam and reduced using Subaru's automated pipeline, which includes a bias subtraction, flat field correction, and a distortion correction.
Suprime-Cam has 10 CCDs which cover a field of view is 34$\arcmin\times27\arcmin$ with a pixel scale of $0.20\arcsec$; each CCD which contains a TNO is analyzed separately.

The TNO images were acquired with exposures of $\le300$ seconds to minimize trailing due to object motion, but carefully accounting for the small amount of motion reduces photometric uncertainty.
The rate of motion of the targets ranged from 1.5--5.1$\arcsec/$hr, with a median rate of 2.5$\arcsec/$hr.
The majority of the targets were sidereally tracked, however due to an error in the program setup, for 2013 UQ$_{15}$ (o3l77), 2013 UL$_{15}$ (o3l43), and 2001 QF$_{331}$ (o3l06PD), the object was tracked instead of the stars.
We use the Trailed Image Photometry in Python (TRIPPy) software package, which makes use of a pill-shaped aperture \citep{fraser2016phot}.
This pill-shaped aperture is an extension of the circular aperture photometric measurement method, where the aperture shape is elongated based on object rate of motion, which is used to make a more accurate aperture correction than a purely circular aperture.
For PSFs derived from sidereal tracked stars, aperture corrections can be determined to better than 0.01 magnitudes for the pill aperture \citep{fraser2016phot}.
The photometry unique to this work is reported in Tables \ref{phot_subaru} and \ref{phot_51}; full tables of the Col-OSSOS photometry will be included in a forthcoming data release paper \citep{colossos}.

\subsection{SDSS Color Calibration}

In order to compare the colors acquired using different bandpass filters on different facilities, it is necessary to properly characterize the flux measurement of each telescope and scale to a common system.
The Col-OSSOS data are scaled to the Sloan Digital Sky Survey (SDSS) Release 13 \citep{SDSS13} magnitudes from photometry of in-frame SDSS catalog stars.
Those SDSS stars were then used to determine the color transform between the SDSS and Gemini filter sets.
From the GMOS photometry, color terms between the GMOS ($r_\mathrm{G}$, $g_\mathrm{G}$, $z_\mathrm{G}$) and SDSS ($r_\mathrm{SDSS}$, $g_\mathrm{SDSS}$, $z_\mathrm{SDSS}$) systems were determined to be:
\begin{equation}
   r_\mathrm{G} = r_\mathrm{SDSS} - (0.060\pm0.03)\times(g_\mathrm{SDSS}-r_\mathrm{SDSS})
\end{equation} 
\begin{equation}
   g_\mathrm{G} = g_\mathrm{SDSS}-(0.140\pm0.03)\times(g_\mathrm{SDSS}-r_\mathrm{SDSS})
\end{equation} 
\begin{equation}
   z_\mathrm{G} = z_\mathrm{SDSS}-(0.027\pm0.018)\times(r_\mathrm{SDSS}-z_\mathrm{SDSS})
\end{equation} 
\noindent
Similar techniques were utilized to measure SDSS-Subaru color terms ($Rc_\mathrm{S}$, $z_\mathrm{S}$, $i_\mathrm{S}$), and calibrate the Subaru observations.
The color terms were determined to be:
\begin{equation}
Rc_\mathrm{S} = Rc_\mathrm{SDSS}-(0.044\pm0.023)\times(g_\mathrm{SDSS}-r_\mathrm{SDSS})
\end{equation} 
\begin{equation}
i_\mathrm{S} = i_\mathrm{SDSS}-(0.096\pm0.019)\times(g_\mathrm{SDSS}-r_\mathrm{SDSS})
\end{equation} 
\begin{equation}
z_\mathrm{S} = z_\mathrm{SDSS}-(0.077\pm0.034)\times(g_\mathrm{SDSS}-r_\mathrm{SDSS})
\end{equation} 
\noindent
An additional color term was used to convert $Rc_\mathrm{SDSS}$ to $r_\mathrm{SDSS}$ based on the multi-band photometry of SDSS sources on frame \citep{jordi2006}.
\begin{equation} Rc_\mathrm{SDSS}-r_\mathrm{SDSS} =  (-0.153 \pm 0.003)\times(r_\mathrm{SDSS}-i_\mathrm{SDSS}) - (0.117 \pm 0.003) 
\end{equation}
\noindent
Photometry from Subaru and Gemini, converted to the SDSS system, are presented in Tables \ref{phot_subaru} and \ref{phot_51}, respectively.

\subsection{Determining Colors from TNO Photometry}

An accurate color determination requires multi-band photometry taken within a short time or carefully corrected to mitigate variation due to light curve and phase effects \citep{duffard2009,fraser2015}.
Variations in target brightness were detected across the Col-OSSOS Gemini imaging sequence.
To approximately account for this, a model in which a TNO exhibits a linear variation in source brightness and constant colors through the $grz$ range was fit to the observations in a least-squares sense.
If the Subaru photometric measurements were taken within $\pm0.5$~hours of an $r$ band measurement from Gemini, the temporally closest $r$ band magnitude was considered sufficiently unaffected by rotational variation and used to calculate the $r-z$ color.
In some cases the observations were not simultaneous, and a Subaru $Rc$ measurement was used to determine the $r-z$ color; additional uncertainty is propagated into the color estimates in Table \ref{tno_color}.
Three of our targets have duplicate $r-z$ colors, each with one acquired from Subaru, and one from Gemini.
In all cases, these color measurements are consistent within their uncertainties.
The consistency of the measurements demonstrates the accuracy of the calibration method.
One $z$ band measurement lacks an associated $r$ or $Rc$ measurement within $\pm0.5$ hours; this object (2014 UL$_{225}$) is included in the photometry Table \ref{phot_subaru} for completeness, but no $r-z$ color is reported as the variations due to the lightcurve are unknown.
Four TNOs were also measured in $i$ band; their $r-i$ colors are consistent with expectation based on their $g-r$ colors \citep{ofek2012}.

\begin{table}[h!]
\setlength{\tabcolsep}{9.pt}
\caption{TNO Photometry Sequences from Subaru in SDSS Magnitudes}
\label{phot_subaru}
\begin{center}
\begin{tabular}{ l c  c  c  c  c  c  c c  c c c c }
\hline\hline
Survey &MPC &Filter & Magnitude & MJD    \\
ID & ID& (SDSS) & &   \\\hline
   o4h50   &   2014 UE$_{225}$   &   $i$   &   22.36 $\pm$ 0.05   &   56894.40706 \\
  o4h50   &   2014 UE$_{225}$   &   $z$   &   22.01 $\pm$ 0.06   &   56894.41911 \\
   o4h50   &   2014 UE$_{225}$   &   $z$   &   22.20 $\pm$ 0.08   &   56897.42265 \\
   o4h50   &   2014 UE$_{225}$   &   $Rc$ ($r$)   &   22.31 $\pm$ 0.04   &   56894.42717 \\ 
\arrayrulecolor{gray}\hline    
  o4h45   &  2014 UD$_{225}$   &   $z$   &   22.56 $\pm$ 0.18   &   56897.42265 \\
\arrayrulecolor{gray}\hline    
  o4h01   &   2014 UJ$_{225}$   &   $z$   &   22.47 $\pm$ 0.14   &   56897.42053 \\
\arrayrulecolor{gray}\hline    
   o4h20   &   2014 UL$_{225}$   &   $z$   &   22.92 $\pm$ 0.14   &   56897.42265 \\
\arrayrulecolor{gray}\hline    
  o4h31   &   2014 UM$_{225}$  &   $z$   &   23.12 $\pm$ 0.17   &   56897.42481 \\
\arrayrulecolor{gray}\hline    
   o3l43   &   2013 UL$_{15}$   &   $i$   &   22.48 $\pm$ 0.05   &   56892.40975 \\
   o3l43   &   2013 UL$_{15}$  &   $z$   &   22.32 $\pm$ 0.07   &   56892.41698 \\
   o3l43   &   2013 UL$_{15}$   &   $Rc$ ($r$)      &   22.71 $\pm$ 0.04   &   56892.42619 \\
\arrayrulecolor{gray}\hline    
 o3l39   &   2016 BP$_{81}$   &   $z$   &   22.32 $\pm$ 0.20   &   56896.41319 \\
 o3l39   &   2016 BP$_{81}$   &   $Rc$ ($r$)      &   22.81 $\pm$ 0.08   &   56896.42616 \\
\arrayrulecolor{gray}\hline    
  o3l63   &   2013 UN$_{15}$  &   $Rc$ ($r$)    &   23.80 $\pm$ 0.09   &  56895.42795  \\
   o3l63   &   2013 UN$_{15}$  &   $z$   &   23.12 $\pm$ 0.15   &   56895.42145 \\
\arrayrulecolor{gray}\hline    
 o3l09   &   2013 US$_{15}$    &   $z$   &   22.46 $\pm$ 0.10   &   56897.41821 \\
\arrayrulecolor{gray}\hline    
  o3l01   &   2013 UR$_{15}$   &   $i$   &   22.70 $\pm$ 0.07   &   56894.40391 \\
   o3l01   &   2013 UR$_{15}$   &   $z$   &   22.48 $\pm$ 0.12   &   56894.41690 \\
\arrayrulecolor{gray}\hline    
  o3l46   &   2013 UP$_{15}$   &   $Rc$ ($r$)    &   23.99 $\pm$ 0.1   &   56896.42408 \\
 o3l46   &   2013 UP$_{15}$   &   $z$   &   23.7 $\pm$ 0.3   &   56896.41111 \\
\arrayrulecolor{gray}\hline    
  o3l06PD   &   2001 QF$_{331}$    &   $i$   &   22.44 $\pm$ 0.04   &   56892.40761 \\
 o3l06PD   &   2001 QF$_{331}$    &   $z$   &   22.15 $\pm$ 0.06   &   56892.41914 \\
\arrayrulecolor{gray}\hline    
  o3l15   &   2013 SZ$_{99}$   &   $z$   &   23.38 $\pm$ 0.2   &   56897.42720 \\
\arrayrulecolor{gray}\hline    
  o3l77   &   2013 UQ$_{15}$   &   $z$   &   22.93 $\pm$ 0.15   &   56897.41603 \\
\arrayrulecolor{gray}\hline    
  o3l79   &   2013 SA$_{100}$   &   $Rc$ ($r$)   &   22.87 $\pm$ 0.05   & 56896.42187   \\
   o3l79   &   2013 SA$_{100}$   &   $z$   &   22.4 $\pm$ 0.1   &  56896.40887  \\
\arrayrulecolor{gray}\hline    
  o3l76   &   2013 SQ$_{99}$   &   $z$   &   22.54 $\pm$ 0.18   &   56896.40887 \\
   o3l76   &   2013 SQ$_{99}$   &  $Rc$ ($r$)    &   23.15 $\pm$ 0.08   &  56896.42187  \\
\arrayrulecolor{black}\hline
\end{tabular}
\end{center}
\footnotesize{All Subaru exposures were 150 seconds.}
\end{table}

\begin{table}[h!]
\setlength{\tabcolsep}{9.pt}
\caption{TNO Photometry Sequences of the 5:1 Resonators from Gemini}
\label{phot_51}
\begin{center}
\begin{tabular}{ l c  c  c  c  c  c  c c  c c c c }
\hline\hline
Survey &MPC &Filter & Magnitude & MJD & Exposure Time & Magnitude & Magnitude  \\
ID & ID&  & (Gemini) &  & (s)& (Gemini)& (SDSS)   \\\hline
L3y02   &   2003 YQ$_{179}$   &   $r$   &   22.91 $\pm$ 0.02   &   57308.63136   &   300  \\
L3y02   &   2003 YQ$_{179}$    &   $r$   &    &    &   &    22.91 $\pm$ 0.02   &  22.95 $\pm$ 0.02 \\
L3y02   &   2003 YQ$_{179}$   &   $g$   &   23.61 $\pm$ 0.04   &   57308.63533   &   300  \\
L3y02   &   2003 YQ$_{179}$     &   $g$   &    &    &   &      23.61 $\pm$ 0.04   &  23.72 $\pm$ 0.04 \\
L3y02   &   2003 YQ$_{179}$   &   $z$   &   22.50 $\pm$ 0.05   &   57308.63933   &   300  \\ 
L3y02   &   2003 YQ$_{179}$     &   $z$   &    &    &   &     22.50 $\pm$ 0.05   &  22.51 $\pm$ 0.05   \\
\arrayrulecolor{gray}\hline
HL7j4   &   2007 LF$_{38}$   &   $r$   &   23.12 $\pm$ 0.02   &   57435.63929   &   200  \\
HL7j4   &   2007 LF$_{38}$   &   $r$   &   23.10 $\pm$ 0.02   &   57435.65007   &   200  \\ 
HL7j4   &   2007 LF$_{38}$      &   $r$   &    &    &   &      23.11 $\pm$ 0.01   &  23.14 $\pm$ 0.02\\
HL7j4   &   2007 LF$_{38}$   &   $g$   &   23.59 $\pm$ 0.03   &   57435.64328   &   300  \\
HL7j4   &   2007 LF$_{38}$     &   $g$   &    &    &   &       23.59 $\pm$ 0.03   &  23.67 $\pm$ 0.03 \\
HL7j4   &   2007 LF$_{38}$   &   $z$   &   22.86 $\pm$ 0.04   &   57435.64728   &   300  \\
HL7j4   &   2007 LF$_{38}$      &   $z$   &    &    &   & 22.86 $\pm$ 0.04   &    22.87 $\pm$ 0.04  \\      
\arrayrulecolor{gray}\hline    
HL7c1   &   2007 FN$_{51}$   &   $r$   &   23.69 $\pm$ 0.03   &   57432.52142   &   300  \\
HL7c1   &   2007 FN$_{51}$   &   $r$   &   23.68 $\pm$ 0.03   &   57432.55252   &   300  \\
HL7c1   &   2007 FN$_{51}$     &   $r$   &    &    &   &       23.68 $\pm$ 0.05   & 23.72 $\pm$ 0.05\\
HL7c1   &   2007 FN$_{51}$   &   $g$   &   24.35 $\pm$ 0.03   &   57432.50984   &   300  \\
HL7c1   &   2007 FN$_{51}$   &   $g$   &   24.25 $\pm$ 0.03   &   57432.51365   &   300  \\
HL7c1   &   2007 FN$_{51}$   &   $g$   &   24.33 $\pm$ 0.03   &   57432.51746   &   300  \\
HL7c1   &   2007 FN$_{51}$   &   $g$   &   24.33 $\pm$ 0.03   &   57432.53317   &   300  \\
HL7c1   &   2007 FN$_{51}$   &   $g$   &   24.42 $\pm$ 0.03   &   57432.53697   &   300  \\
HL7c1   &   2007 FN$_{51}$   &   $g$   &   24.39 $\pm$ 0.04   &   57432.54078   &   300  \\
HL7c1   &   2007 FN$_{51}$   &   $g$   &   24.37 $\pm$ 0.04   &   57432.55649   &   300  \\
HL7c1   &   2007 FN$_{51}$   &   $g$   &   24.29 $\pm$ 0.04   &   57432.56030   &   300  \\
HL7c1   &   2007 FN$_{51}$   &   $g$   &   24.41 $\pm$ 0.05   &   57432.56410   &   300  \\
HL7c1   &   2007 FN$_{51}$      &   $g$   &    &    &   &             24.35 $\pm$ 0.05   &  24.45 $\pm$ 0.05\\
HL7c1   &   2007 FN$_{51}$   &   $z$   &   23.06 $\pm$ 0.04   &   57432.52536   &   300  \\
HL7c1   &   2007 FN$_{51}$   &   $z$   &   23.11 $\pm$ 0.04   &   57432.52916   &   300  \\
HL7c1   &   2007 FN$_{51}$   &   $z$   &   23.25 $\pm$ 0.05   &   57432.54477   &   300  \\
HL7c1   &   2007 FN$_{51}$   &   $z$   &   23.41 $\pm$ 0.07   &   57432.54858   &   300  \\
HL7c1   &   2007 FN$_{51}$      &   $z$   &    &    &   &   23.18 $\pm$ 0.05   & 23.19 $\pm$ 0.05  \\
\arrayrulecolor{black}\hline
\end{tabular}
\end{center}
\end{table}

\begin{table}[h!]
\setlength{\tabcolsep}{9.5pt}
\caption{TNO Colors (SDSS)}
\label{tno_color}
\begin{center}
\begin{tabular}{ l c  c  c  c  c  c  c c  c c c c }
\hline\hline

Survey& MPC  & $g-r$ & $r-z$    & $r-z$ & $r-i$    \\
 ID& ID & Gemini & Gemini  & Subaru & Subaru \\\hline 
o4h50   &    2014 UE$_{225}$    &   1.02 $\pm$ 0.01   &   ...       &   0.30 $\pm$ 0.07 &   -0.05 $\pm$ 0.09 \\ 
o4h01   &    2014 UJ$_{225}$    &   0.65 $\pm$ 0.02   &   ...       &   0.69 $\pm$ 0.14&   ...   \\
o4h45  &    2014 UD$_{225}$   &   0.69 $\pm$ 0.02   &   ...      &   0.59 $\pm$ 0.19 &   ...  \\
o4h31   &    2014 UM$_{225}$   &   0.80 $\pm$ 0.03   &   ...      &   0.48 $\pm$ 0.17 &   ...  \\ 
o3l43   &    2013 UL$_{15}$    &   0.91 $\pm$ 0.04   &   ...      &   0.39 $\pm$ 0.08 &   0.23 $\pm$ 0.06 \\  
o3l39  &   2016 BP$_{81}$    &   0.55 $\pm$ 0.02   &   ...      &   0.5 $\pm$ 0.2&   ...  \\
o3l77   &    2013 UQ$_{15}$    &   0.54 $\pm$ 0.02   &   ...      &   0.26 $\pm$ 0.16 &   ... \\
o3l63 (2014B)   &    2013 UN$_{15}$   &   1.05 $\pm$ 0.04   &   0.38 $\pm$ 0.09     &   0.52 $\pm$ 0.2&   ...  \\
o3l63 (2015B)  &    2013 UN$_{15}$   &   ...   &   0.73 $\pm$ 0.07     &   ... &   ...  \\
o3l09   &    2013 US$_{15}$     &   1.05 $\pm$ 0.01   &   ...      &   0.81 $\pm$ 0.1&   ...  \\
o3l01   &    2013 UR$_{15}$    &   0.66 $\pm$ 0.05   &   ...     &   0.85 $\pm$ 0.13 &   0.58 $\pm$ 0.07 \\
o3l46   &    2013 UP$_{15}$    &   0.90 $\pm$ 0.01   &   ...      &   0.41 $\pm$ 0.3&   ... \\
o3l06   &    2001 QF$_{331}$     &   0.87 $\pm$ 0.02   &   ...       &   0.75 $\pm$ 0.06 &   0.46 $\pm$ 0.05 \\
o3l15   &    2013 SZ$_{99}$    &   0.70 $\pm$ 0.02   &   ...       &   0.37 $\pm$ 0.2 &   ...  \\
o3l79 (2014B)  &    2013 SA$_{100}$    &   0.61 $\pm$ 0.01   &   0.47 $\pm$ 0.01      &   0.47 $\pm$ 0.11&   ... \\  
o3l79 (2015B)   &    2013 SA$_{100}$    &   0.66 $\pm$ 0.01   &   0.41 $\pm$ 0.01      & ... &   ... \\
o3l76  &    2013 SQ$_{99}$    &   0.99 $\pm$ 0.02   &   0.54 $\pm$ 0.03     &   0.62 $\pm$ 0.18&   ...  \\   
o4h19   &    2014 UK$_{225}$    &   0.96 $\pm$ 0.02   &   0.70 $\pm$ 0.02    &   ... &   ...  \\
o4h29   &    2014 UH$_{225}$    &   0.55 $\pm$ 0.02   &   0.36 $\pm$ 0.06       &   ...  &   ... \\
o5t09PD   &    2014 UA$_{225}$    &   0.91 $\pm$ 0.03   &   0.69 $\pm$ 0.01     &   ... &   ...  \\
o5s06  &    2015 RW$_{245}$    &   0.71 $\pm$ 0.01   &   0.41 $\pm$ 0.05    &   ...&   ...   \\
o5t04   &    2015 RU$_{245}$    &   0.81 $\pm$ 0.01   &   0.59 $\pm$ 0.02     &   ...  &   ... \\
o5t11PD   &    2001 QE$_{298}$    &   0.87 $\pm$ 0.01   &   0.60 $\pm$ 0.03   &   ... &   ...  \\
o5t31  &    2015 RT$_{245}$    &   0.91 $\pm$ 0.03   &   0.60 $\pm$ 0.01    &   ... &   ...  \\
o5s05   &    2015 RV$_{245}$    &   0.61 $\pm$ 0.04   &   0.41 $\pm$ 0.05     &   ... &   ...  \\
o5s16PD   &    2004 PB$_{112}$    &   0.74 $\pm$ 0.01   &   0.58 $\pm$ 0.01      &   ... &   ...  \\
L3y02  & 2003 YQ$_{179}$  &  0.77 $\pm$ 0.05   &   0.44 $\pm$ 0.05   &   ... &   ...  \\
HL7j4 & 2007 LF$_{38}$ &  0.53 $\pm$ 0.04   &   0.27 $\pm$ 0.05    &   ...&   ...   \\
HL7c1 & 2007 FN$_{51}$ &  0.73 $\pm$ 0.03   &   0.53 $\pm$ 0.03    &   ...  &   ... \\
\hline
\end{tabular}
\end{center}
\end{table}

\section{Results}

\begin{figure}[h!]
\begin{center}
\makebox[\textwidth][c]{\includegraphics[width=0.6\textwidth]{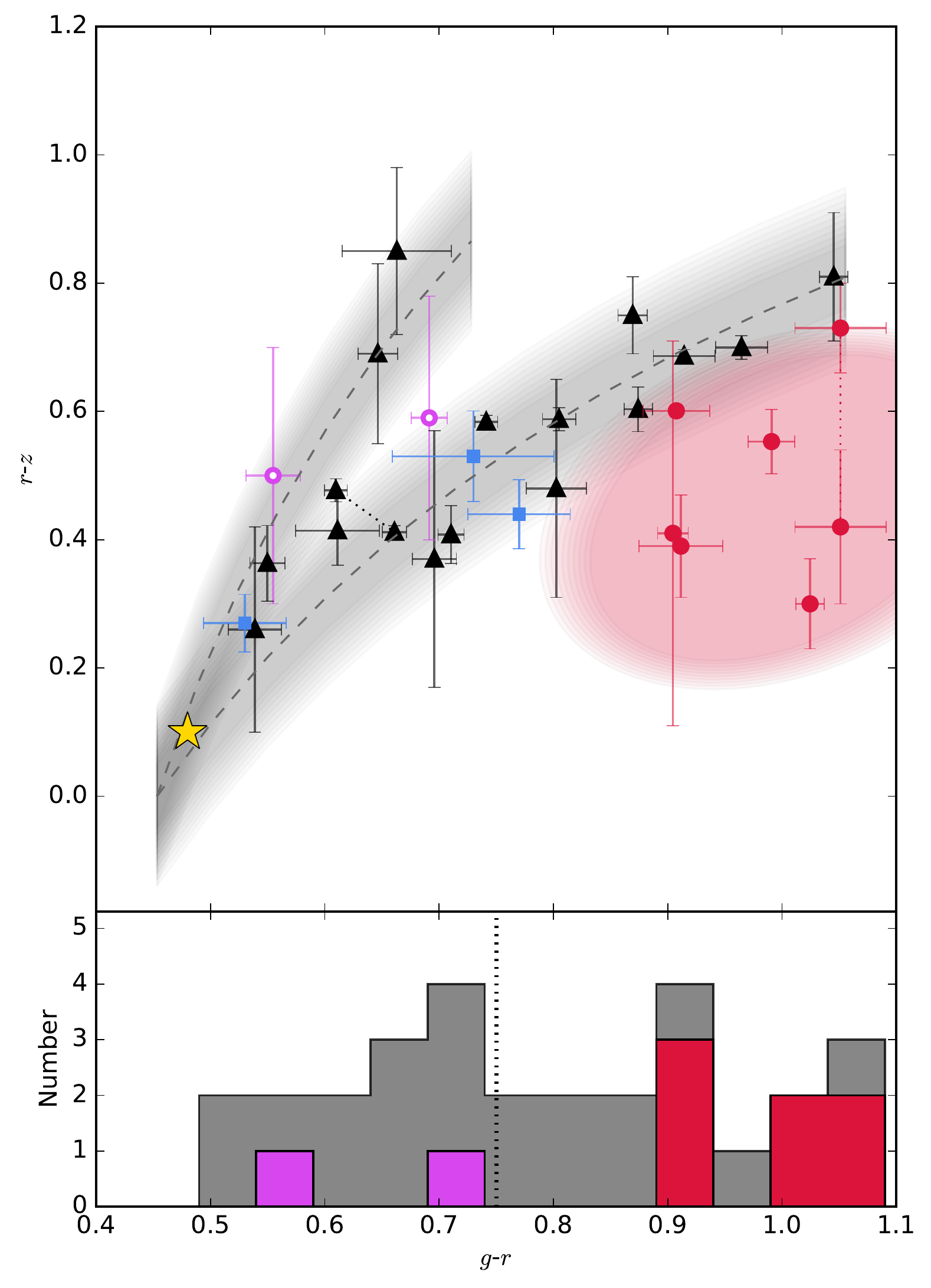}}
\caption{Upper: All targets with $z$ band photometry, identified as dynamically excited TNOs (black triangles), 5:1 resonators (blue squares), or cold classical TNOs (solid red and empty pink circles).  In the case of measurements from both Subaru and Gemini in the same semester, a weighted mean is shown.  For multiple measurements in different semesters, both are shown connected with a dotted line.  Two surface models based on \cite[dashed lines]{fraser2012} for TNO surfaces are plotted and are well matched by the two dynamically excited TNO surfaces.  The `blue binary' cold classical TNOs with neutral surfaces are indicated with pink empty circles, these follow the same color trends as the excited objects.  Solar colors are indicated by the star.  The cold classical objects clearly occupy a separate clump.  The gray arcs and red ellipse are a visual guide to indicate the approximate regions occupied by the three surfaces classes; the width of the arcs is similar to \citet{fraser2012}, but in different photometric bands.  Lower: Histogram of $g-r$ color values for these TNOs with approximate separation from the literature at $g-r\sim$0.75 (dotted line).  Dynamically excited objects (including 5:1 resonators) are shown in gray, cold classicals are red and pink (blue binaries).  The histogram does not correctly separate the two dynamically excited surface types.}
\label{color1}
\end{center}
\end{figure}

In Figure \ref{color1}, we present the measured colors of the TNOs in $g-r$ and $r-z$.
The $z$ band measurements increase the distinction between different surface types suggested by the $g$ and $r$ band measurements for the TNOs.
The $r-z$ colors of the objects span a large range of values, $\sim$0.7 magnitudes.
In Figure \ref{color1}, the $r-z$ versus $g-r$ structure indicating different surface reflectivity becomes apparent.
The commonly reported $g-r$ bi-modality dividing surface color types at $g-r\sim0.75$ noted in previous studies \citep[e.g.][]{peixinho2015} is indicated in the histogram.
Given the size of our sample, unsurprisingly, we do not find statistically significant evidence of bi-modality.
As this topic is discussed quite thoroughly in other works with samples more appropriate to test the existence of bi-modality, we adopt the accepted conclusion that the excited populations possess two compositional classes as evidenced by their bimodal optical colors \citep{tegler1998,peixinho2003,peixinho,peixinho2015,tegler03,tegler2016,doressoundiram2007,fraser2012,wong2017}.
However, Figure \ref{color1} indicates that dividing these objects requires more than a simple optical color cut; we use a surface model based on \cite{fraser2012} to describe the color range occupied by two dynamically excited TNOs compositional classes.
Because of their albedos, cold classical objects are expected to have different surface properties \citep{brucker2009,vilenius2014,lacerda2014}.
The cold classical objects form a third surface type identifiable in $g-r$ and $r-z$.

The dynamically excited TNOs have both a neutral and a red surface group in $g-r$.
The `neutral' objects in $g-r$ ($g-r\lesssim$0.75) show a roughly linear trend of increasing $r-z$ colors with increasing $g-r$ colors.
The Spearman rank test \citep{spearman1904} finds that these colors are correlated ($\rho$=0.82) with 3$\sigma$ significance.
The red dynamically excited TNOs have a different trend of increasing $r-z$ colors with increasing $g-r$ colors at 2$\sigma$ significance, but with a shallower slope ($\rho$=0.72).
With $g-r$ colors alone, it is unclear if the dynamically excited TNOs with $0.6\lesssim g-r \lesssim 0.8$ belong to the red or neutral surface groups.
Due to the clear correlation of $g-r$ and $r-z$, however, the difference in surface colors of these two classes becomes obvious; those objects with red $r-z$ are neutral class members.

A model of the two types of dynamically excited TNO surfaces is presented in Figure \ref{color1}.
We modeled the approximate range of color occupied by a TNO surface class after the approach of \cite{fraser2012}: a simple Hapke surface model \citep{hapke2002} defines the overall reflectivity of a mix of two materials with different surface reflectivity.
The \cite{fraser2012} geometric mixing model uses different reflectivity of two materials in each of the filters; these materials combine in varying amounts to reproduce the possible range of color occupied by a TNO surface class.
Only three materials, one common between the two populations, are all that is necessary to reproduce the colors of the two surface types.
The precise surface reflectivity used in the models is not informative because only the relative reflectivity in the different bands affects the material's color; as a result we describe only the comparative reflectivity as in \cite{fraser2012}.
The Hubble Space Telescope Wide Field Camera 3 filters used in the modeling by \cite{fraser2012} are sufficiently different from the filter selection here that this new data set provides independent confirmation of the published models.

We calculated model reflectivities for the three surface components that well represent the $g-r$ and $r-z$ surface colors of our TNO sample.
Similar to the \cite{fraser2012} model, we find that the red and neutral surface types can be approximately accounted for by two different red-end components, but a common neutral-end component.
Our data imply that this neutral component is a roughly neutral reflector through $grz$, though it could be slightly less reflective in $z$.
The $z$ band reflectivity was not explored in \cite{fraser2012}, but this lower $z$ band reflectance is consistent with their speculation that the neutral material is silicates.
Our two surface models have different red components, which combine with the neutral/blue component in different ratios to produce the range of colors of each of the two dynamically excited spectral classes.
The models account for the range and distribution of $grz$ colors of the dynamically excited objects quite effectively.

The cold classical TNOs occupy a different range of $g-r$ and $r-z$ space than do the dynamically excited objects; they are red in $g-r$ and less reflective in $z$ band.
The objects identified in Figure \ref{color1} as cold classical TNOs were selected from the classical sub-population based on pericenters $>$38 AU, semi-major axes 39.4--48.0 AU, and inclinations $<$6$^{\circ}$.
Photometry in $g$, $r$, and $z$ band demonstrates a clear difference between the red cold classical surfaces and dynamically excited TNO surfaces.
The Spearman rank test finds that the $r-z$ and $g-r$ colors of red cold classical objects do not show a statistically significant correlation; this may be due to the small sample size or the underlying color distribution.
The cold classicals also include 2013 UL$_{15}$, which shows clear variation in $r-z$ between the two observing epochs; though the color variation is large, both measurements fall within the surface colors of cold classicals.
(Although TNOs typically do not display variability in $g-r$, variability has been identified beyond $\sim0.9\micron$ \citep{fraser2015}.)
As it is unclear if the cold classicals have correlated colors or if they clump in $grz$ space, we do not model these surfaces using a geometric mixing model.
In our sample of cold classicals, the only exceptions to the unique red cold classical surfaces are those 2 objects identified as belonging to the recently identified class of blue binary objects on cold classical orbits \citep{fraser2017}.
These blue binary TNOs have colors consistent with the dynamically excited population in all bands studied here: $g$, $r$, and $z$, which supports the theory that these objects formed inward of their current location and were pushed outward and hence share a primordial origin with the dynamically excited TNOs \citep{fraser2017}.

 \begin{figure}[h!]
\begin{center}
\makebox[\textwidth][c]{\includegraphics[width=0.6\textwidth]{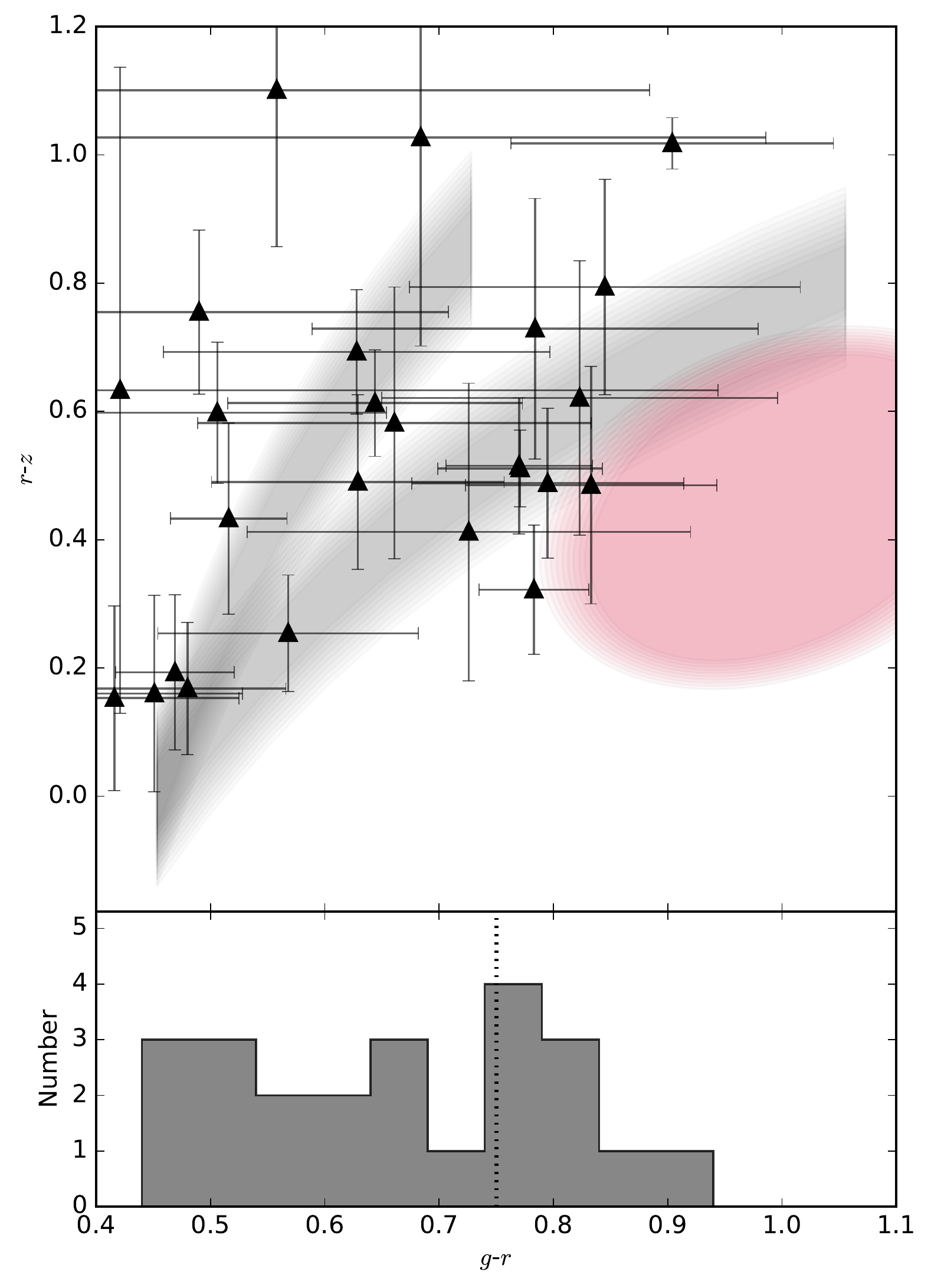}}
\caption{The data from \cite{ofek2012}, presented identically to Figure \ref{color1}.  Upper:  All TNOs are dynamically excited objects.  Because of the large photometric uncertainties and additional uncertainty resulting from non-simultaneous color measurements, the TNOs do not clearly separate into two surface classes (grey shaded regions).  However, the colors are representative of the extent of the dynamically excited object colors.  Crucially, the overlap with the cold classical surface region (red) is minimal.  Lower: Histogram of $g-r$ color values for \cite{ofek2012} TNOs.}
\label{color2}
\end{center}
\end{figure}

Previous work by \cite{ofek2012} provides further evidence that the red cold classical objects occupy a distinct region of $g-r$ vs. $r-z$ color space.
The photometry of TNOs from the SDSS presented by \cite{ofek2012} in which $g$, $r$, and $z$ photometry are available is presented in Figure \ref{color2}.
Due to the depth of the SDSS, this sample is entirely dynamically excited TNOs.
The larger uncertainties in the photometry and the non-simultaneous color measurements obscure the distinction between the two dynamically excited surface classes, but the range of colors is representative of the dynamically excited objects as a whole.
The surface reflectance of the dynamically excited TNOs measured in \cite{ofek2012} do not extend into the region we have identified as cold classical TNO surfaces, confirming that these surface colors are unique to cold classical objects.

We verify the unique surface properties of cold classical objects by using the Kolmogorov-Smirnov (KS) test \citep{ks2D}.
We compared the sample of dynamically excited objects to the dynamically cold objects using both the one and two dimensional KS test. 
To ensure a fair comparison, we only consider objects with $g-r>0.85$, the color range occupied by the dynamically selected sample of cold classical objects. 
Bootstrapping random simulated samples was used to calibrate both the 1D and 2D KS test results.
From the 2D KS test applied to the $g-r$ and $r-z$ colors, we find that the probability that the red dynamically hot and cold TNOs share the same parent distribution is only 2\%. 
Similarly, the 1D KS test on the $r-z$ distribution reports only a 1\% probability that the two samples share the same parent color distribution. 
It appears that the cold classical objects possess a different $grz$ color distribution than do the dynamically excited objects. 
Our findings support the assertion drawn from the high albedos of cold classicals \citep{brucker2009} and the analysis of non-simultaneous colors of TNOs in the Hyper Suprime-Cam Subaru Strategic Program which revealed indications of distinct cold classical surfaces \citep{terai2017}; the cold classical objects possess a unique surface type.

In the wavelengths studied in \cite{fraser2012}, the red cold classical colors overlapped significantly with the red dynamically excited surfaces, as is the case in $g-r$ and $g-i$ in other previous studies.
The $grz$ surface reflectance of red cold classicals and red excited objects provides a new diagnostic for identifying interlopers in the cold classical region and red cold classical surfaces elsewhere in the Kuiper belt.
In $grz$, the red cold classical objects clearly exhibit a different compositional class than the excited TNOs, in agreement with the high albedos exhibited by these objects compared to the duller excited TNOs \citep{brucker2009,vilenius2014}.

\section{Discussion}

The $g$, $r$, and $z$ band photometry show that these TNOs have three surface types.
The TNO colors are consistent with the known bi-modality in $g-r$ \citep[e.g.][]{tegler1998,peixinho2003,peixinho,peixinho2015,tegler03,tegler2016,fraser2012,wong2017}.
The addition of the $z$ band measurement makes it possible to separate TNOs where the $g-r$ surface groups overlap and determine which surface group the TNOs belong to: neutral excited, red excited, red cold classical.
The red cold classical surfaces occupy a distinct region of the $grz$ color space, and the two neutral cold classicals (blue binaries) are consistent with the dynamically excited objects.
The neutral and red excited TNOs show two different correlated slopes between the $g-r$ and $r-z$ colors.
This reddening is well represented by the two component geometric composition model \citep{fraser2012}.

The variation in $z$ band color is indicative of TNO surface properties, and several materials are speculated to be present on TNO surfaces.
`Tholin' is an organic compound which has been reddened through irradiation \citep{roush2004}; a material of this type is typically thought to be responsible for the red spectral slopes of TNOs in $g$, $r$, and $i$ band which should extend with the same slope through $z$ band as well.
However, if TNO surfaces include contributions from an iron-rich material, such as olivine or pyroxene \citep{usgs}, these materials are less reflective in $z$ band.
The inclusion of a silicate material in the mixing model, such olivine or pyroxene material, would result in a range of $z$ band reflectivity.
A silicate component for TNO composition was a good match for the neutral component of the TNO surface models from \cite{fraser2012}, and this is better demonstrated in the models in Figure \ref{color1} where the neutral component in $g$ and $r$ has a reduced reflectance in $z$ band.
We speculate that cold classicals have surfaces richer in silicates than the red excited objects or perhaps a different surface silicate material. 
A larger sample of precise multi-band photometry or spectroscopy searching for silicate absorption could further constrain the TNO surface components.

The three 5:1 resonators in our sample have a range of surface colors in $g-r$ and $r-z$.
Two of the objects are consistent with the surface model of red excited objects, and the third is consistent with a neutral excited TNO surface; none of the 5:1 resonators resemble the cold classical object surfaces.
The known 5:1 resonators have surfaces consistent with the dynamically excited population, which implies the dynamically excited and distant resonant objects share the same source population.
Dynamically excited populations display a range of $g-r$ surface colors, seen here and in previous work \citep[e.g.][]{tegler1998}.
This range of surface colors may have resulted from formation in different locations closer to the Sun \citep{brown}, followed by scattering into the outer Solar System.
\cite{pike2015} speculates that the 5:1 objects are captured from the scattering objects, and the range of 5:1 resonator surface colors is consistent with capture from the dynamically excited scattering object colors.

We find that $z$ band photometry provides a powerful tool to more precisely discriminate between three different surface groups and clearly identifies the red cold classical TNO surfaces as unique in the Kuiper belt.
These data show that when TNO colors overlap in $g-r$, $z$ band can be used to effectively divide the TNOs into three surface classifications: red cold classical TNOs, dynamically excited red TNOs, and dynamically excited neutral TNOs.
TNOs are sufficiently bright in $z$ band for this measurement to be a reasonable addition to a TNO color survey.
Expanding the use of $z$ band photometry would provide a useful tool for tracing the dynamical history of the region, as it enables the identification of cold classical surfaces outside the classical belt as well as the identification of hot classical object interlopers on cold classical orbits.

\acknowledgements

Based on observations obtained at the Gemini Observatory, acquired through the Gemini Observatory Archive, and processed using the Gemini IRAF package, which is operated by the Association of Universities for Research in Astronomy, Inc., under a cooperative agreement with the NSF on behalf of the Gemini partnership: the National Science Foundation (United States), the National Research Council (Canada), CONICYT (Chile), Ministerio de Ciencia, Tecnolog\'{i}a e Innovaci\'{o}n Productiva (Argentina), and Minist\'{e}rio da Ci\^{e}ncia, Tecnologia e Inova\c{c}\~{a}o (Brazil).
Based on data collected at Subaru Telescope, which is operated by the National Astronomical Observatory of Japan.
The authors recognize and acknowledge the very significant cultural role and reverence that the summit of Maunakea has always had within the indigenous Hawaiian community. We are fortunate to have the opportunity to conduct observations from this mountain. 
MB acknowledges support from UK STFC grant ST/L000709/1.
MES was supported by Gemini Observatory.

\software{
{Astropy v-1.3}
{TRIPPy v-0.5.1}
{SciPy v-0.17.1}
{Matplotlib v-1.5.1}
	       }

\bibliographystyle{/Applications/TeX/aastexv6.0/aasjournal}

\end{document}